\documentclass{article}

\usepackage{hyperref}
\hypersetup{
    colorlinks=true,
    linkcolor=blue,     
    citecolor=violet, 
    filecolor=magenta,  
    urlcolor=cyan,      
}
\usepackage{cite} 
\usepackage{booktabs}  
\usepackage{ltablex}   
\usepackage{multirow}  
\usepackage{makecell}

\usepackage{amssymb,amsfonts,amsmath}
\usepackage{cite,enumerate,float}
\usepackage{color}
\usepackage{tikz}
\usetikzlibrary{arrows,snakes,backgrounds}
\usepackage{url}
\usepackage[vcentermath]{youngtab}

\usepackage{relsize}

\usepackage{enumitem,varwidth}

\def\be{\begin{eqnarray}}
\def\ee{\end{eqnarray}}
\def\nn{\nonumber}

\def\ve{\varepsilon}
\def\vz{\vec{z}}
\def\vy{\vec{y}}
\def\vrho{\vec{\rho}}

\def\va{\vec{\alpha}}

\definecolor{red}{rgb}{1,0,0}
\definecolor{orange}{rgb}{1,0.5,0}
\definecolor{violet}{rgb}{0.7,0,1}

\hoffset= - 1.0in         

\begin{document}

\title{\vspace{1.5cm}\bf
Baker-Akhiezer functions for \\ Koornwinder-van Diejen integrable system
}

\author{
Liudmila Bishler
}

\date{ }

\maketitle

\vspace{-5.5cm}

\begin{center}
  \hfill FIAN/TD-20/26\\

\end{center}

\vspace{3.5cm}

\begin{center}

bishlerlv@lebedev.ru

{\small {\it Lebedev Physical Institute, Moscow 119991, Russia}}\\

\end{center}

\vspace{.1cm}

\begin{abstract}

The Koornwinder-van Diejen integrable system is a counterpart of the trigonometric Ruijsenaars integrable system (associated with the  root system $A_n$) for the root system $(C_n^\vee, C_n)$. It depends on the standard Macdonald parameters $q$ and $t$ as well as the Koornwinder (Askey-Wilson) parameters $a$, $b$, $c$, $d$. Different specializations of the parameters connect it to the other classical root systems from the $(C_n^\vee, C_n)$ family.

The eigenfunctions of the Koornwinder-van Diejen integrable system are the symmetric Koornwinder polynomials; for certain values of the parameters $t$, $a$, $b$, $c$, $d$ they coincide with the sum over the Weyl group of the Baker-Akhiezer functions introduced by O.A.\,Chalykh for the classical root system $BC_n$. The main idea behind this paper is to elucidate Chalykh's formalism and apply it to other affine root systems. We define Baker-Akhiezer functions for all root systems in the $(C_n^\vee, C_n)$ family. We also present some explicit expressions for Baker-Akhiezer functions.

\end{abstract}

\bigskip

\newcommand\smallpar[1]{
  \noindent $\bullet$ \textbf{#1}
}

\section{Introduction}

The Koornwinder-van Diejen integrable system consists of $n$ commuting Hamiltonians, which were constructed by T.\,Koornwinder \cite{Koorn} and J.F.\,van Diejen \cite{vD}. It depends on the standard Macdonald parameters $q$ and $t$ as well as the parameters $a$, $b$, $c$, $d$ and is symmetric under the permutations of $a$, $b$, $c$, $d$. It emerged when Koornwinder looked for a multi-variable analog of the Askey-Wilson polynomials \cite{AW} and generalized several Macdonald polynomials corresponding to the classical root systems \cite{Mac,MacConj} (except for the root system $A_n$) into one construction which is called Koornwinder polynomial. He found the first Koornwinder Hamiltonian, which was later extended to the system of commuting Hamiltonians by van Diejen. Koornwinder polynomials are the Laurent polynomials in variables $x_1, \dots, x_n$, they are symmetric under the action of the hyperoctahedral group, so symmetric under the permutation of the variables as well as the inversions $x\to x^{-1}$. They also depend on the parameters $q$ and $t$ and $a$, $b$, $c$, $d$.

The Koornwinder-van Diejen integrable system corresponds to the affine root system $(C_n^{\vee},C_n)$  (in Macdonald notation \cite{Macr}) which we denote as $C^{\vee}C$ for brevity. The 
specializations of the Koornwinder parameters allow to get the integrable systems associated with the other classical affine root systems: reduced $B_n$, $C_n$, $D_n$, $B_n^{\vee}$, $C_n^{\vee}$ and $BC_n$ and non-reduced $(BC_n,C_n)$, $(C_n^{\vee},BC_n)$, $(B_n^{\vee},B_n)$. The parameters are listed in Table \ref{tab:root_systems} below. 

\begin{table}[ht]
\centering
\caption{Koornwinder parameters}
\label{tab:root_systems}
\begin{tabular}{l c l}
\toprule
\textbf{} & \textbf{Root system} & \textbf{Parameters} \\
\midrule
\multirow{6}{*}{\textbf{Reduced}} 
    & $B_n$       & $(a,b,c,d)=(-q,-1,q,t_s^2)$ \\
    & $C_n$       & $(a,b,c,d)=(-t_l,-qt_l,t_l,qt_l)$ \\
    & $D_n$       & $(a,b,c,d)=(-q,-1,q,1)$ \\
    & $B_n^\vee$  & $(a,b,c,d)=(-q,-t_l,q,t_l)$ \\
    & $C_n^\vee$  & $(a,b,c,d)=(-t_s,-1,q,qt_s)$ \\
    & $BC_n$      & $(a,b,c,d)=(-qt_l,-1,qt_l,t_s^2)$ \\
\midrule
\multirow{3}{*}{\textbf{Non-reduced}} 
    & $(BC_n,C_n)$     & $(a,b,c,d)=(-t_s,-qt_l,t_l,qt_l)$ \\
    & $(C_n^\vee,BC_n)$ & $(a,b,c,d)=(-t_s,-qt_s,q,t_l)$ \\
    & $(B_n^\vee,B_n)$  & $(a,b,c,d)=(-q,-t_s,q,t_l)$ \\
\bottomrule
\end{tabular}
\end{table}

I.\,Cherednik developed another approach to the  Macdonald polynomials with the double affine Hecke algebra (DAHA) \cite{CherednikDAHA, Cherednik}. He found the more elementary integrable system which consists of Cherednik operators. Their eigenfunctions are the non-symmetric Macdonald polynomials \cite{Opdam,MacNS,ChNS}, which yield their symmetric counterparts upon the weighted symmetrization. The non-symmetric Koornwinder system is described by $C^{\vee}C$ DAHA \cite{Noumi,Sahi,Sahi2,St,Chalykh,ChDunkl,CR}. 
Within the framework of DAHA the non-symmetric Koornwinder polynomials can be constructed recursively with the analog of the Knop-Sahi recurrence \cite{KS,HHL}. 

The Koornwinder integrable system is the counterpart \cite{BMP} of the trigonometric Ruijsenaars integrable system \cite{RS,RS2} which corresponds to the root system $A_n$. It can be described by a vertical ray $(0,1)$ of the Ding-Iohara-Miki (DIM) \cite{DI,Miki}  algebra and plays a profound role in the description of the integrable systems within DIM. It's connected to DAHA of type $A_n$ due to the correspondence of DIM and spherical DAHA \cite{dFK1}.

The eigenfunctions of the trigonometric Ruijsenaars integrable system form the Macdonald triad \cite{MMP3, dFK2}:  Noumi-Shiraishi power series \cite{NS},  Macdonald polynomials and at $t = q^{-m}$ --- Baker-Akhiezer (BA) functions \cite{Ch2002}.  Macdonald polynomials and BA functions are two polynomial reductions of Noumi-Shiraishi power series. The ambitious plan to construct a triad for the general $C^{\vee}C$ case seems unlikely to be realized \cite{BMP}; however, analogs of the triad can exist at special cases. In this paper we closely consider one element of these potential triads --- Baker-Akhiezer functions.

O.A.\,Chalykh \cite{Ch2002} considered the Macdonald operators at
\begin{equation}\label{Chparm1}
    t_{\alpha} = q^{-m_{\alpha}}, \quad m_{\alpha} \in \mathbb{Z}_{\geq 0}
\end{equation}
as well as the first Koornwinder Hamiltonian at{\footnote{Note that there are additional conditions on these parameters, we consider them in section \ref{sec:Chalykhformalism}.}} 
\begin{equation}\label{Chparm2}
    a=-q^{-m_1},\,\, b=-q^{-m_2},\,\, c=q^{-m_3},\,\, d=q^{-m_4},\,\, t=q^{-m}, \quad m_1,m_2, m_3,m_4,m \in \mathbb{Z}_{\geq 0}
\end{equation}
and found non-symmetric eigenfunctions of other type: Baker-Akhiezer functions.  Chalykh  formulates his approach for finite reduced and irreducible root systems and non-reduced root system $BC_n$.
He studied the poles of the corresponding Hamiltonians and formulated the conditions on their eigenfunctions that cancel those poles out. He showed that these conditions are linear equations. What is more these linear equations with the right base function uniquely define eigenfunctions of corresponding Hamiltonians. The special place of the parameters (\ref{Chparm1}) and (\ref{Chparm2}) is also evident when we consider Macdonald and Koornwinder polynomials: at these parameters some coefficients in front of monomials go to zero and the initial polynomial splits into a sum of Chalykh's Baker-Akhiezer functions with the permuted variables.

In this paper we consider Chalykh's approach for the sets of the parameters $a$, $b$, $c$, $d$ corresponding to different affine root systems from Table \ref{tab:root_systems}.  We go into the details on how Chalykh's approach works for these  affine root systems and calculate some BA functions. There are three components in Chalykh's formalism: a base function, which consists of monomials corresponding to positive roots. A summation domain which defines the number of coefficients. And finally equations. They determine BA function uniquely up to a normalization coefficient. 

\bigskip

The paper is organized as follows. In section \ref{KvDintsystem} we give an overview of the Koornwinder-van Diejen integrable system. There are two known types of eigenfunctions of this system: symmetric Koornwinder polynomials and Baker-Akhiezer functions. This paper is devoted to Baker-Akhiezer functions. And in section \ref{sec:Chalykhformalism} we consider Chalykh's definition of Baker-Akhiezer functions for finite reduced and irreducible root systems and non-reduced and irreducible root system $BC_n$. In section \ref{sec:Results} we show how Chalykh's approach works for other affine root systems in $C^{\vee}C$ family. We summarize different cases in Table \ref{tab:multirow} and provide a more detailed description of the rank one case in section \ref{sec:Examples}. 
In section \ref{sec:D2case} we consider the first non-trivial case for root system $D_n$ ($D_1$ case is trivial) and present the BA function for root system $D_2$. Section \ref{Concl} is a conclusion. 
\bigskip

\paragraph{Notation.} 
Note that we use the parameters $q^2$ and $t^2$ compared to Chalykh's original notation. The parameters in Table \ref{tab:root_systems} are also written for $q^2$, $t_s^2$, $t_l^2$. In the case of the root system $B_n$ we use the parameter $t_s^2$ instead of $t_s$ (in comparison with the classifications from \cite{BMP,Yanagida}) in order to align two ways to get BA functions for this system: via  Chalykh's standard formalism for the reduced root systems and Chalykh's formalism for the root system $BC_n$.

The notation $x^y$ for $n$ variables $x_i$, $y_i$, $i=1,\ldots,n$ means
\be
x^y:=\prod_{i=1}^n x_i^{y_i}
\ee
and similarly $x^{\vec y}$.

We define the quantum numbers
$$
[x]_q:=\frac{q^x-1}{q-1}
$$
and also
$$
\{x\}:=x-x^{-1}.
$$

\section{Koornwinder-van Diejen integrable system \label{KvDintsystem}}

 The first Koornwinder Hamiltonian was constructed by Koornwinder \cite{Koorn} (note that parameters $q$ and $t$ are squared as compared with the original Koornwinder notation):
\be\label{KH}
H^K_1(x_1, \dots, x_n):=\sum_{i=1}^nt^{2(1-n)}{(1-ax_i)(1-bx_i)(1-cx_i)(1-dx_i)\over \eta(1-x_i^2)(1-q^2x_i^2)}\prod_{j\ne i}
{(1-t^2x_ix_j)(1-t^2x_i/x_j)\over (1-x_ix_j)(1-x_i/x_j)}\left(q^{2\hat D_i}-1\right)+\nn\\
+\sum_{i=1}^nt^{2(1-n)}{(1-ax_i^{-1})(1-bx_i^{-1})(1-cx_i^{-1})(1-dx_i^{-1})\over \eta(1-x_i^{-2})(1-q^2x_i^{-2})}\prod_{j\ne i}
{(1-t^2x_i^{-1}x_j^{-1})(1-t^2x_j/x_i)\over (1-x_i^{-1}x_j^{-1})(1-x_j/x_i)}\left(q^{-2\hat D_i}-1\right), 
\ee
where $\eta = \frac{\sqrt{a b c d}}{q}$, $\hat D_i  = x_i \frac{\partial}{\partial x_i} $. 

In the simplest rank one case (Askey-Wilson case) the system depends on just one variable $x$ and the first Hamiltonian is the following

\be\label{KH1}
H^K_1(x):={(1-ax)(1-bx)(1-cx)(1-dx)\over \eta(1-x^2)(1-q^2x^2)}\left(q^{2\hat D}-1\right)+{(1-ax^{-1})(1-bx^{-1})(1-cx^{-1})(1-dx^{-1})\over \eta(1-x^{-2})(1-q^2x^{-2})}\left(q^{-2\hat D}-1\right).
\ee

The first Koornwinder Hamiltonian is embedded into the system of $n$ commuting Hamiltonians constructed by van Diejen \cite{vD}:
\begin{equation}
    \begin{aligned}
        H_r^{K} = \sum_{\substack{J \subset \{1,\dots,n\},\, |J| = r \\ \varepsilon_j = \pm 1,\, j\in J}} \,\, \sum_{\substack{\emptyset \subsetneqq J_1 \subsetneqq \dots \subsetneqq J_s = J \\ 1 \leq s \leq r}} (-1)^{s-1} \prod_{1\leq s'\leq s} h_{\varepsilon(J_{s'} \setminus J_{s'-1});J_{s'}^C} (q^{2\sum_{j\in J_1}\varepsilon_j \hat D_j } -1).
    \end{aligned}
\end{equation}
The first sum runs over all subsets of $\{1,\dots,n\}$ with $r$ elements and over all sign reversals $\varepsilon = \{+1,-1\}$. The second sum runs over all strictly increasing sequences of subsets in $J$.

$J^C$ is a complement of $J$ with respect to $\{1,\dots,n\}$,
\begin{equation}
    \begin{aligned}
    h_{\varepsilon J;K} := \prod_{j\in J} \frac{1}{\eta}{(1-ax_j^{\varepsilon_j})(1-bx_j^{\varepsilon_j})(1-cx_j^{\varepsilon_j})(1-dx_j^{\varepsilon_j})\over (1-x_j^{2\varepsilon_j})(1-q^2 x_j^{2\varepsilon_j})} \times \prod_{\substack{j,j' \in J \\ j < j'}} \frac{1}{t^2} \frac{(1-t^2 x_j^{\varepsilon_j} x_{j'}^{\varepsilon_{j'}})}{(1- x_j^{\varepsilon_j} x_{j'}^{\varepsilon_{j'}})} \frac{(1- q^2 t^2 x_j^{\varepsilon_j} x_{j'}^{\varepsilon_{j'}})}{(1- q^2 x_j^{\varepsilon_j} x_{j'}^{\varepsilon_{j'}})} \times  \\
    \times  \prod_{\substack{j \in J \\ k \in K}} \frac{1}{t^2} \frac{(1-t^2 x_j^{\varepsilon_j} x_{k})}{(1- x_j^{\varepsilon_j} x_{k})} \frac{(1-t^2 x_j^{\varepsilon_j}/ x_{k})}{(1- x_j^{\varepsilon_j}/ x_{k})}.
     \end{aligned}
\end{equation}

As an integrable system it is connected with $C^{\vee} C$ DAHA: The first Koornwinder Hamiltonian coincides with the Weyl-symmetric combination of the Cherednik Hamiltonians, when acting on the Weyl-symmetric functions:
\be
H^K_1=\sum_{i=1}^n(C_i+C_i^{-1})-[n]_t\Big(t^{n-1}\eta+{1\over t^{n-1}\eta}\Big)\Big|_{symm}.
\ee

\section{Chalykh's formalism \label{sec:Chalykhformalism}}

O.A.\,Chalykh \cite{Ch2002,ChEt} constructed eigenfunctions for the Macdonald operators and the first Koornwinder Hamiltonian for the special values of parameters. These eigenfunctions are enumerated not with Young diagrams (as Macdonald polynomials), but with a set of arbitrary complex numbers $\lambda$. 

The idea behind Chalykh's formalism is to study poles of a corresponding Hamiltonian and to formulate the conditions on its eigenfunctions that cancel those poles out. The conditions are linear equations which with a right base function uniquely define eigenfunctions of corresponding Hamiltonians.

First we describe Chalykh's formalism for the root systems $R = A_n,B_n,C_n,D_n$ as well as for the exceptional root systems $E_6, E_7,E_8$, $F_4$ and $G_2$. In this case the equations come from poles of the corresponding Macdonald operators. For the root system $A_n$ these operators coincide with Ruijsenaars Hamiltonians. For detailed description of Chalykh's approach in case of the root system $A_n$ see \cite{MMPapproach}.

Apart from reduced and irreducible root systems Chalykh also considers non-reduced finite root system $BC_n$. In this case Chalykh's formalism is built around the Koornwinder Hamiltonian (\ref{KH}). We consider it in the second part of this section.

In contrast with Chalykh's paper \cite{Ch2002}, we interchanged the space and the cospace in the definitions of base functions (\ref{base1}), (\ref{base2}) and periodicity conditions (\ref{ChEq}) and (\ref{eqa})-(\ref{eq2})  so that the resulting BA functions would match the corresponding Macdonald and Koornwinder polynomials from Macdonald's definition \cite{Mac}. There are some explicit expressions in \cite{Feigin,AS,Noumi2}. The other difference is that Chalykh's equations are written for $q^2$.

\subsection{Reduced root systems}

Let $R$ be a reduced and irreducible root system, $W_R$ --- its Weyl group, $R_{+}$ --- a set of positive roots. For each orbit of the Weyl group there are different $m_{\alpha} \in \mathbb{Z}_{\geq 0}$:
\begin{equation}
    t_{\alpha} = q^{-m_{\alpha}}.
\end{equation}

The base function is the sum with arbitrary coefficients $\psi_{\{k_{\alpha}\}}$ of monomials $q^{k_{\alpha}(\alpha|z)}$ in variables $x_i = q^{z_i}$. The summation goes up to $k_{\alpha} = m_{\alpha}$ and $\rho_m$ is used to shift the summation domain and make it symmetric with respect to zero:  
\begin{equation}
    \Psi_{\{m_{\alpha}\} } (\vec x, \vec y;\, q^2) = q^{(\lambda+\rho_m|\,z)} \sum_{\{k_{\alpha}\}} \psi_{\{k_{\alpha}\}} q^{-\sum_{\vec\alpha\in R_{+}} k_{\alpha} (\alpha|z)},
    \label{base1}
\end{equation}
where $k_{\alpha} = 0,\dots m_{\alpha}$ for each positive root $\alpha\in R_{+}$, $x_i = q^{z_i}$ and $y_i = q^{\lambda_i}$, $(x|y)$ denotes the scalar product between $\vec x$ and $\vec y$,
\begin{equation}
    \vrho_{m} = \frac{1}{2}\sum_{\alpha \in R_{+}} m_{\alpha} \va.
\end{equation}
The base function (\ref{base1}) satisfies the equations (for each $\alpha \in R_{+}$)
\begin{equation}
    \Psi_{\{m_{\alpha}\}}\left(\vz+j\va^{\vee},\vy;\,q^2 \right)=\Psi_{\{m_{\alpha}\}}\left(\vz-j\va^{\vee},\vy;\,q^2\right)\quad \text{for} \quad  j=1,\dots m_{\alpha} \quad \text{at} \quad q^{(\alpha|z)} = 1.
    \label{ChEq}
\end{equation}

\subsection{Non-reduced finite root system $BC_n$} 

The Koornwinder BA function is defined for the following choice of parameters $a,b, c, d$:
\begin{equation}
 t=q^{-m}, \quad   a = - q^{-m_1}, \quad b = - q^{-m_2},\quad  c = q^{-m_3},\quad  d = q^{-m_4},
 \label{abcdM}
\end{equation}
where $m, m_1, m_2, m_3, m_4 \in \mathbb{Z}_{\geq 0}$ and the following conditions are satisfied:
\begin{equation}
    m_1+m_2+1 = 2M_1 \quad \text{and} \quad  m_3+m_4+1 = 2M_2, \quad M_1, M_2 \in \mathbb{Z}_{\geq 0}.
  \label{ChKooCond}
\end{equation}
\noindent
Even though Koornwinder polynomials are symmetric under the permutation of parameters $a,b, c, d$, we still chose them with different signs and it separates them. What is more according to (\ref{ChKooCond}) they should have different ``parities'' inside each pair, for example:

\begin{equation}
    a = -q^{-(2m_a+1)}, \quad b = - q^{-2m_b},\quad  c = q^{-(2m_c+1)},\quad  d = q^{-2m_d},
    \label{abcd}
\end{equation}
where all $m_a,m_b, m_c, m_d \in \mathbb{Z}_{\geq 0}$. 

The root system $BC_n$ consists of the roots of the root systems $B_n$ and $C_n$, so there are short $\ve_i$ and long $2\ve_i$ roots as well as standard roots $\ve_i\pm \ve_j$. However, it is sufficient to use only the  roots of the root system $B_n$ to construct the BA base function for the root system $BC_n$  (see \cite{Koorn,Ch2002}). The positive roots of $B_n$ are the following 
\begin{equation}
    R_+^{B_n} = \left\{\ve_i | 1\leq i\leq n \right\} \cup \left\{ \ve_i\pm \ve_j| 1\leq i < j \leq n\right\}, 
    \end{equation}
so there are two parameters $m_{\alpha}$ for the short and standard roots. We denote them $m_{\ve}$ and $m_{\ve\pm\ve}$ respectively:
\begin{align}
   & m_{\ve} = M_1+M_2 = 1+\frac{1}{2}(m_1+m_2+m_3+m_4), \label{eq:shortrange}\\
   & m_{\ve\pm\ve} = m. 
\end{align}
The definition of the Weyl vector is the same as in the previous section and for the parameters $m_{\ve}$ and $m_{\ve\pm\ve}$ it is the following
\begin{equation}
    (\vec\rho_m)_i=m(n-i)+{M_1+M_2\over 2}.
\end{equation}
The Koornwinder BA base function has the same structure (\ref{base1}), but with a particular set of monomials:
\begin{equation}
    \begin{aligned}
      & \Psi^{K}_{m,m_1,m_2,m_3,m_4}(\vec x,\vec y ; \,q^2) =  \\ & = \prod_{i=1}^n x_i^{{\lambda_i}+m(n-i)+{M_1+M_2\over 2}} \sum_{\{k_i\}=0}^{M_1+M_2}\sum_{\{l_{ij},l'_{ij}\}=0}^{m}\psi_{\{k_i,l_{ij},l'_{ij}\}}
     \left( \prod_{i=1}^n\left(\frac{1}{x_i}\right)^{k_i}\prod_{1\leq i<j\leq n} \left(\frac{x_j}{x_i} \right)^{l_{ij}}\left(\frac{1}{x_i x_j} \right)^{l'_{ij}}  \right).
     \label{base2} 
    \end{aligned}
\end{equation}
The equations are written for denominators of the Hamiltonian (\ref{KH}). For the parameters (\ref{abcd}) the periodicity conditions take the form (for $i=1,\dots,n$):
\begin{equation}
    \Psi^{K}_{m,m_1,m_2,m_3,m_4} (x_i q^{s}, \vec y;\, q^2) =   \Psi^{K}_{m,m_1,m_2,m_3,m_4} (x_i q^{-s}, \vec y;\, q^2),
    \label{BigEq}
\end{equation}
where
\begin{alignat}{2}
    s &= 1,3\dots, 2m_a+1 && \quad\quad\quad \text{at } x_i = -1, \label{eqa} \\
    s &= 2,4\dots, 2m_b   && \quad\quad\quad \text{at } x_i = -1, \label{eqb} \\
    s &= 1,3\dots, 2m_c+1 && \quad\quad\quad \text{at } x_i = 1,  \label{eqc} \\
    s &= 2,4\dots, 2m_d   && \quad\quad\quad \text{at } x_i = 1,  \label{eqd}
\end{alignat}
and we also have the equations:
\begin{align}
     & \Psi^{K}_{m,m_1, m_2, m_3, m_4} (x_i q^{2s}, \vec y;\, q^2) =   \Psi^{K}_{m,m_1, m_2, m_3, m_4} (x_j q^{2s}, \vec y;\, q^2), \quad s = 1, \dots, m \quad \quad &&\text{at} \quad x_i = x_j, \\
    & \Psi^{K}_{m,m_1, m_2, m_3, m_4} (x_i q^{2s}, \vec y;\, q^2) =   \Psi^{K}_{m,m_1, m_2, m_3, m_4} (x_j q^{-2s}, \vec y;\, q^2), \quad s = 1, \dots, m \quad \quad &&\text{at} \quad x_i = x_j^{-1}. 
\end{align}
The equations (\ref{eqa})-(\ref{eqd}) are written for parameters (\ref{abcdM}) when $m_1, m_2, m_3, m_4 \in \mathbb{Z}_{\geq 0}$. However, specifications of the parameters from Table~\ref{tab:root_systems} that correspond to other root systems do not satisfy these conditions. Chalykh's formalism can still be applied, but with modifications, in particular some of the shifts (\ref{eqa})-(\ref{eqb}) for the equation (\ref{BigEq}) vanish. In the next section we sum up the definitions of BA functions associated with different sets of parameters from Table~\ref{tab:root_systems}.

\section{BA function for root systems in $C^{\vee}C$ family  \label{sec:Results}}

The essential ingredients in the definition of BA function are the specification of parameters $a$, $b$, $c$, $d$ via $q$, $t_s$, $t_l$ (where $t_s$ and $t_l$ denote the parameters associated with short and long roots correspondingly). There are also conditions on $t_s$ and $t_l$ essential for existence of BA function, a range of summation for short roots $N$ and a set of shifts $s$ in the equations corresponding to short roots. 

The base function in this case is the same 
\begin{equation}
    \begin{aligned}
      & \Psi^{K}_{\{m\}}(\vec x,\vec y ; \,q^2) =  \\ & = \prod_{i=1}^n x_i^{{\lambda_i}+m(n-i)+\rho} \sum_{\{k_i\}=0}^{N}\sum_{\{l_{ij},l'_{ij}\}=0}^{m}\psi_{\{k_i,l_{ij},l'_{ij}\}}
     \left( \prod_{i=1}^n\left(\frac{1}{X_i}\right)^{k_i}\prod_{1\leq i<j\leq n} \left(\frac{x_j}{x_i} \right)^{l_{ij}}\left(\frac{1}{x_i x_j} \right)^{l'_{ij}}  \right), 
     \label{base} 
    \end{aligned}
\end{equation}
where $\{m\}$ denotes a list of parameters $m_{\alpha}$ for each orbit of the Weyl group of the corresponding root system.
$\rho$, $N$ and $X$ are listed in Table \ref{tab:multirow} as well as shifts $s$ and conditions for the equation:

\begin{equation}
    \Psi^{K}_{\{m\}} (x_i q^{s}, \vec y;\, q^2) =   \Psi^{K}_{\{m\}} (x_i q^{-s}, \vec y;\, q^2).
    \label{BigEq1}
\end{equation}
BA function also satisfies the equations:
\begin{align}
     & \Psi^{K}_{\{m\}} (x_i q^{2s}, \vec y;\, q^2) =   \Psi^{K}_{\{m\}} (x_j q^{2s}, \vec y;\, q^2), \quad s = 1, \dots, m \quad \quad &&\text{at} \quad x_i = x_j, \label{eq1} \\
    & \Psi^{K}_{\{m\}} (x_i q^{2s}, \vec y;\, q^2) =   \Psi^{K}_{\{m\}} (x_j q^{-2s}, \vec y;\, q^2), \quad s = 1, \dots, m \quad \quad &&\text{at} \quad x_i = x_j^{-1}. \label{eq2}
\end{align}

Askey-Wilson Baker-Akhiezer (AW BA) function associated with root system $R$ has a form 
\begin{equation}
    \Psi_{\{m\}}^{AW} (x,y;\,q^2) = x^{\lambda +\rho} \sum_{l=0}^{N} \psi_l \left( \frac{1}{X}\right)^l, 
 \end{equation}
and satisfies the equations 
\begin{equation}
     \Psi_{\{m\}}^{AW}(xq^{s},y;\,q^2) = \Psi_{\{m\}}^{AW}(xq^{-s},y;\,q^2).
     \label{SmallEq}
\end{equation}
for a set of $s$ at $x=\pm 1$ from Table \ref{tab:multirow}. 

\begingroup
\setlength{\aboverulesep}{8pt}
\setlength{\belowrulesep}{8pt}
\begin{tabularx}{\textwidth}{c|X|X|X}
\caption{Key components of BA function definition} \label{tab:multirow} \\
\toprule
\makecell{\textbf{Root} \\ \textbf{system}} & \makecell{\textbf{Parameters} \\ $a$, $b$, $c$, $d$ } & \makecell{\textbf{Base function} \\ \textbf{components}} &  \makecell{\textbf{Shifts}} \\
\midrule
\endfirsthead

\multicolumn{4}{c}{Table \ref{tab:multirow} \textit{(continued})} \\
\toprule
\makecell{\textbf{Root} \\ \textbf{system}} & \makecell{\textbf{Parameters} \\ $a$, $b$, $c$, $d$ } & \makecell{\textbf{Base function} \\ \textbf{components}} &  \makecell{\textbf{Shifts}} \\
\midrule
\endhead

\bottomrule
\endlastfoot

\makecell[c]{$(C^{\vee},C)$} & \makecell[l]{$a$, $b$, $c$, $d$\\[4pt]
 $a = -q^{-m_1}$,   $b = -q^{-m_2}$, \\ $  c = q^{-m_3}$,  $ d = q^{-m_4}$ \\[4pt]
 $m_1+m_2+1 = 2M_1$ \\
 $m_3+m_4+1 = 2M_2$ \\
 $M_1, M_2 \in \mathbb{Z}_{\geq 0}$}
 & \makecell[l]{
 $X = x$, \\
 $N = M_1+M_2$  \\ 
 $\rho = N/2$ 
 }
 & \makecell[l]{
  $s = 1, 3, \dots, m_1 \quad \quad \text{at} \quad x = -1$, \\
  $s = 2, 4, \dots, m_2 \quad \quad \text{at} \quad x = -1$, \\
  $s = 1, 3, \dots, m_3 \quad \quad \text{at} \quad x = 1$, \\
  $s = 2, 4, \dots, m_4 \quad \quad \text{at} \quad x = 1$
}  \\
\midrule
\makecell[c]{$B$} & \makecell[l]{$-q$, $-1$, $q$, $t_s^2$ \\[4pt]
 $t_s = q^{-m_s}$ \\[4pt]
 $m_s\in \mathbb{Z}_{\geq 0}$} 
 & \makecell[l]{
 $X = x$, \\
 $N = m_s$, \\
 $\rho = N/2$ 
 }
 & \makecell[l]{
  $s=2,4,\dots,2m_s \quad \text{at} \quad x = 1$
}  \\

\midrule
\makecell[c]{$C$} & \makecell[l]{$-t_l$, $-q t_l$, $t_l$, $q t_l$\\[4pt]
 $t_l = q^{-m_l}$ \\[4pt]
 $m_l\in \mathbb{Z}_{\geq 0}$} 
 & \makecell[l]{
 $X = x^2$, \\
 $N = m_l$, \\
 $\rho = N$ 
 }
 & \makecell[l]{
  $s=1,2,\dots,m_l \quad \text{at} \quad x^2 = 1$
}  \\

\midrule
\makecell[c]{$B^{\vee}$} & \makecell[l]{$-q$, $- t_l$, $q$, $t_l$\\[4pt]
 $t_l = q^{- m_l}$  \\[4pt]
 $m_l=0,2,\dots$ (only even)} 
 & \makecell[l]{
 $X = x^2$, \\
 $N = m_l/2$, \\
 $\rho = N$ 
 }
 & \makecell[l]{
  $s=2, 4,\dots,m_l \quad \text{at} \quad x^2 = 1$
}  \\

\midrule
\makecell[c]{$C^{\vee}$} & \makecell[l]{ $-t_s$,$-1$, $q$, $q t_s$\\[4pt]
 $t_s = q^{-m_s}$ \\[4pt]
 $m_s = 1,3,\dots$ (only odd) } 
 & \makecell[l]{
 $X = x$, \\
 $N = m_s$, \\
 $\rho = N/2$ 
 }
 & \makecell[l]{
  $s=2,4\dots,m_s-1 \quad \text{at} \quad x = 1$\\
  $s=1,3\dots,m_s \quad \text{at} \quad x = -1$
}  \\

\midrule
\makecell[c]{$BC$} & \makecell[l]{$-q t_l$, $-1$, $q t_l$, $t_s^2$\\[4pt]
 $t_s = q^{-m_s}$, $t_l = q^{-m_l}$\\[4pt]
$m_s \in \mathbb{Z}_{\geq 0}$,\\ [4pt]
$m_l = 0,2,4,\dots$ (only even)}
 & \makecell[l]{
 $X = x$, \\
 $N = m_s+m_l$, \\
 $\rho = N/2$ 
 }
 & \makecell[l]{
  $s=2,4\dots,2m_s \quad \text{at} \quad x = 1$,\\
  $s=1,3\dots,m_l-1 \quad \text{at} \quad x^2 = 1$
}  \\

\midrule
\makecell[c]{$(BC,C)$} & \makecell[l]{$-t_s$, $-q t_l$, $t_l$, $q t_l$\\[4pt]
 $t_s = q^{-m_s}$, \\
 $t_l = q^{-m_l}$, \\[4pt]
$m_s,m_l = 0,2,4\dots$ or \\
$m_s,m_l = 1,3,5\dots$ }
 & \makecell[l]{
 $X = x$, \\
 $N = (m_s+3m_l)/2$, \\
 $\rho = N/2$ 
 }
 & \makecell[l]{
  for odd $m_s,m_l$: \\
  $  s = 1, 3, \dots m_s \quad \quad \text{at} \quad x = -1 $,\\
  $  s = 2, 4, \dots m_l-1 \quad \quad \text{at} \quad x^2 = 1$,\\
  $  s = 1, 3, \dots m_l \quad \quad \text{at} \quad x = 1$, \\[4pt]
  for even $m_s,m_l$: \\
  $  s = 1, 3, \dots m_l-1 \quad \quad \text{at} \quad x^2 = 1$,\\
  $  s = 2, 4, \dots m_s \quad \quad \text{at} \quad x = -1$,\\
  $  s = 2,4, \dots m_l \quad \quad \text{at} \quad x = 1$
}  \\

\midrule
\makecell[c]{$(C^{\vee},BC)$} & \makecell[l]{$-t_s$, $-q t_s$, $q$, $t_l$\\[4pt]
 $t_s = q^{-m_s}$, \\
 $t_l = q^{-m_l}$, \\[4pt]
$m_l = 0,2,4\dots$ \\
$m_s \in \mathbb{Z}_{\geq 0}$ }
 & \makecell[l]{
 $X = x$, \\
 $N = m_s+m_l/2$, \\
 $\rho = N/2$ 
 }
 & \makecell[l]{
  $  s = 1, 2, \dots m_s \quad \quad \text{at} \quad x = -1 $,\\
  $  s = 2, 4, \dots m_l \quad \quad \text{at} \quad x = 1$
}  \\

\midrule
\makecell[c]{$(B^{\vee},B)$} & \makecell[l]{$-q$, $-t_s$, $q$, $t_l$\\[4pt]
 $t_s = q^{-m_s}$, \\
 $t_l = q^{-m_l}$, \\[4pt]
$m_s,m_l = 0,2,4\dots$}
 & \makecell[l]{
 $X = x$, \\
 $N = (m_s+m_l)/2$, \\
 $\rho = N/2$ 
 }
 & \makecell[l]{
  $  s = 2, 4, \dots m_s \quad \quad \text{at} \quad x = -1 $,\\
  $  s = 2, 4, \dots m_l \quad \quad \text{at} \quad x = 1$
}  \\

\end{tabularx}
\endgroup

In case of root system $D_n$ there are no equations associated with parameters $a$, $b$, $c$, $d$ and there is no parameters except  $t=q^{-m}$. That is why we omitted it from the table. We consider $D_2$ case in the section~\ref{sec:D2case}.

\section{Askey-Wilson Baker-Akhiezer functions \label{sec:Examples}}
The one-variable Koornwinder polynomial does not depend on the parameter $t$, and it is in fact Askey-Wilson polynomial \cite{AW}, manifestly given by the formula \cite{Noumi2}
\be\label{AWP}
K_{[n]}(x)={(ab,ac,ad;q^2)_n\over a^n(q^{2(n-1)}abcd;q^2)_n}\cdot \sum_{k=0}^n{(q^{-2n},q^{2(n-1)}abcd,ax,ax^{-1};q^2)_k\over (q^2,ab,ac,ad;q^2)_k}q^{2k},
\ee
where $n$ is integer and 
\begin{align}
    (x;q^2)_n =  \prod_{j=0}^{n-1} \left ( x^{1/2}q^j-x^{-1/2}q^{-j}\right).
\end{align}

\subsection{The most general case: affine root system $(C^{\vee}_1,C_1)$}
\begin{equation}
    a = -q^{-m_1}, \quad  b = -q^{-m_2}, \quad  c = q^{-m_3}, \quad  d = q^{-m_4},
\end{equation}
where
\begin{equation}
    m_1+m_2+1 = 2M_1, \quad m_3+m_4+1 = 2M_2, \quad\quad M_1, M_2 \in \mathbb{Z}_{\geq 0},
\end{equation}
\begin{equation}
    M_1+M_2 = 1+\frac{1}{2}(m_1+m_2+m_3+m_4).
\end{equation}
Since in each pair there should be one odd and one even number we will agree that $m_1, m_3$ are odd and  $m_2, m_4$ are even. 

Base function 
\begin{equation}
    \Psi^{(C^{\vee}_1,C_1)}_{m_1, m_2, m_3, m_4}(x,y;\,q^2) = x^{\lambda+1/2(M_1+M_2)} \sum_{l=0}^{M_1+M_2} \psi_l \left( \frac{1}{x} \right)^l
\end{equation}
satisfies the equations 
\begin{equation}
     \Psi^{(C^{\vee}_1,C_1)}_{m_1, m_2, m_3, m_4}(xq^{j},y;\,q^2) = \Psi^{(C^{\vee}_1,C_1)}_{m_1, m_2, m_3, m_4}(xq^{-j},y;\,q^2) 
\end{equation}
for 
\begin{align}
  &  j = 1, 3, \dots m_1 \quad \quad \text{at} \quad x = -1,\\
  &  j = 2, 4, \dots m_2 \quad \quad \text{at} \quad x = -1,\\
  &  j = 1, 3, \dots m_3 \quad \quad \text{at} \quad x = 1,\\
  &  j = 2, 4, \dots m_4 \quad \quad \text{at} \quad x = 1.
\end{align}

If $c = -a$ and $d = -b$ one gets $C$-type BA function with base function that depends on $m_1, m_2$ and $M_1 = 1/2(m_1+m_2+1)$:
\begin{equation}
    \Psi^{C^{\vee}C_1}_{m_1, m_2}(x,y;\,q^2) = x^{\lambda+M_1} \sum_{l=0}^{M_1} \psi_l \left( \frac{1}{x^2} \right)^l.
\end{equation}
It satisfies the equations 
\begin{equation}
     \Psi^{C^{\vee}C_1}_{m_1, m_2}(xq^{j},y;\,q^2) = \Psi^{C^{\vee}C_1}_{m_1, m_2}(xq^{-j},y;\,q^2) 
\end{equation}
for 
\begin{align}
  &  j = 1, 3, \dots m_1 \quad \quad \text{at} \quad x^2 = 1,\\
  &  j = 2, 4, \dots m_2 \quad \quad \text{at} \quad x^2 = 1.
\end{align}

\subsection{Root system $B_1$}
BA function that corresponds to root system $B_n$ can be obtained from the Chalykh's standard approach since $B_n$ is a reduced and irreducible finite root system. It can also be described in the framework of the Koornwinder BA function. Koornwinder parameters for $B_n$ are $(a,b,c,d)=(-q,-1,q,t_s^2)$. These two directions lead to the same base function and the same equations and in rank-one case give a factorized solution. 

Positive roots are the following
\begin{equation}
    R_+^{B_n} = \left\{\ve_i | 1\leq i\leq n \right\} \cup \left\{ \ve_i\pm \ve_j| 1\leq i < j \leq n\right\}. 
    \end{equation}

In case of root system $B_1$ we only have the short root and the parameter:
\begin{equation}
   t_s = q^{-m_s},
\end{equation}
then the Weyl vector
\begin{equation}
    \rho_{m} = \frac{1}{2} m_s
\end{equation}
and the base function
\begin{equation}
    \Psi_{m_s} (x,y;\,q^2) = x^{m_s/2 + \lambda/2} \sum_{l=0}^{m_s} \psi_l \left( \frac{1}{x} \right)^l.
\end{equation}
Equations from  Chalykh's formalism for the Macdonald operators (\ref{ChEq}):
\begin{equation}
     \Psi_{m_s} (x q^{2j},y;\,q^2) = \Psi_{m_s} (x q^{-2j},y;\,q^2) \quad \text{for} \quad j=1,\dots,m_s \quad \text{at} \quad x = 1.
     \label{eqB}
\end{equation}
On the other hand, one can start with the Koornwinder Hamiltonian:
\be
H^{K(B_1)}_1(x):=
{(1-t_s^2 x)\over \eta(1-x)} \left(q^{2\hat D}-1\right)+{(1-t_s^2 x^{-1})\over \eta(1-x^{-1})} \left(q^{-2\hat D}-1\right),
\ee
where there is just one pole at $x=1$ which corresponds to (\ref{eqd}), it is exactly the equation (\ref{eqB}).
\bigskip

In this case BA function has a nice solution
\be
\Psi_{m_s}^{B_1}(x,y;q^2)=x^{\lambda+m_s\over 2} \sum_{k=0}^{m_s} q^{2k\lambda} x^{-k} \prod_{j=1}^k{[2(m_s-j+1)]_q[2(j-\lambda-m_s-1)]_q\over[2j]_q[2(\lambda-j)]_q},
\ee
where $[x]_q = \frac{q^x-1}{q-1}$.

We want to point out that since the parameter $\lambda$ is arbitrary, we can choose it to be aligned with $\rho$. We take  $\lambda/2$ for $B_1$ and $C_1^{\vee}$ and $\lambda$ for other root systems.

\subsection{Root system $C_1$}
The positive roots are the following
\begin{equation}
    R_+^{C_n} = \left\{2\ve_i | 1\leq i\leq n \right\} \cup \left\{ \ve_i\pm \ve_j| 1\leq i < j \leq n\right\}.
    \end{equation}
The Koornwinder parameters for $C_n$ are $(a,b,c,d)=(-t_l,-qt_l,t_l,qt_l)$. In the case of $C_1$:
\begin{equation}
   t_l = q^{-m_l},  \,\,\rho_{m} = m_l
\end{equation}
\begin{equation}
    \Psi_{m_l}^{C_1} (x,y;\,q^2) = x^{m_l + \lambda} \sum_{l=0}^{m_l} \psi_l \left( \frac{1}{x^2} \right)^l.
\end{equation}

Equations from  Chalykh's formalism for the Macdonald operators (\ref{ChEq}):
\begin{equation}
     \Psi_{m_l}^{C_1} (x q^{j},y;\,q^2) = \Psi_{m_l}^{C_1} (x q^{-j},y;\,q^2) \quad \text{for} \quad j=1,\dots,m_l \quad \text{at} \quad x^2 = 1
     \label{eqC}
\end{equation}
On the other hand, one can start with the Koornwinder Hamiltonian:
\be
H^{K(C_1)}_1(x):=
{(1-t_l^2 x^2)(1-q^2 t_l^2 x^2)\over \eta(1-x^2)(1-q^2 x^2)} \left(q^{2\hat D}-1\right)+{(1-t_l^2 x^{-2})(1-q^2 t_l^2 x^{-2})\over \eta(1-x^{-2})(1-q^2 x^{-2})} \left(q^{-2\hat D}-1\right).
\ee
All the equations (\ref{eqa})-(\ref{eqd}) play a role here and give  the exact equation (\ref{eqC}).

The solution is the following:
\be
\Psi_{m_l}^{C_1}(x,y;\,q^2)=x^{m_l+\lambda} \sum_{k=0}^{m_l} q^{2k\lambda} x^{-2k} \prod_{j=1}^k{[2(m_l-j+1)]_q[2(j-\lambda-m_l-1)]_q\over[2j]_q[2(\lambda-j)]_q} = \Psi_{m_l}^{B_1}(x^2,y;\,q^2).
\ee

\subsection{Root system $D_1$} 

In this case BA function is trivial. $D_2$ case is in the section \ref{sec:D2case}. 

\subsection{Affine root system $B_1^{\vee}$}

The Koornwinder parameters for $B_n^{\vee}$ are $(a,b,c,d)=(-q,-t_l,q,t_l)$.   
The Koornwinder Hamiltonian in this case:
\be
H^{K(B_1^{\vee})}_1(x):={(1-t_l^2 x^2)\over \eta(1-x^2)} \left(q^{2\hat D}-1\right)+{(1-t_l^2 x^{-2})\over \eta(1-x^{-2})} \left(q^{-2\hat D}-1\right).
\ee
There are poles at $x^2=1$ (eq.  (\ref{eqb}), (\ref{eqd})) which leaves us with the equations 
\begin{equation}
     \Psi_{m_l}^{B_1^{\vee}} (x q^{j},y;\,q^2) = \Psi_{m_l}^{B_1^{\vee}} (x q^{-j},y;\,q^2) \quad \text{for} \quad j=2,4,\dots,2m_l \quad \text{at} \quad x^2 = 1.
     \label{eqBvee}
\end{equation}
The parameters
\begin{equation}
    t_l = q^{-2m_l} \quad \text{for} \quad m_{l} \in \mathbb{Z}_{\geq 0}, 
\end{equation}
the base function:
\begin{equation}
    \Psi_{m_l}^{B_1^{\vee}} (x,y;\,q^2) = x^{m_l + \lambda} \sum_{l=0}^{m_l} \psi_l \left( \frac{1}{x^2} \right)^l.
\end{equation}
And the general solution:
\be
\Psi_{m_l}^{B^{\vee}_1}(x,y;\,q^2)=x^{m_l+\lambda} \sum_{k=0}^{m_l} q^{4k\lambda} x^{-2k} \prod_{j=1}^k{[4(m_l-j+1)]_q[4(j-\lambda-m_l-1)]_q\over[4j]_q[4(\lambda-j)]_q} = \Psi_{m_l}^{B_1}(x^2,y;\,q^4).
\ee

\subsection{Affine root system $C_1^{\vee}$}

The Koornwinder parameters for $C_n^{\vee}$ are $(a,b,c,d)=(-1,-t_s,q t_s,q)$  and 
\begin{equation}
    t_s = q^{-m_s}, \quad m_s =2m_b+1 =  1, 3, \dots  \quad \text{(only for odd values of the parameter)}
\end{equation}
The Koornwinder Hamiltonian in this case:
\be
H^{K(C_1^{\vee})}_1(x):={(1+t_s x)(1-q t_s x)\over \eta(1-x)(1+q x)}\left(q^{2\hat D}-1\right)+{(1+t_s x^{-1})(1-q t_s x^{-1})\over \eta(1-x^{-1})(1+q x^{-1})}\left(q^{-2\hat D}-1\right).
\ee
This time we are left with  eqs. (\ref{eqa}) and (\ref{eqd}): 
\begin{equation}
     \Psi_{m_s}^{C_1^{\vee}} (x q^{j},y;\,q^2) = \Psi_{m_s}^{C_1^{\vee}} (x q^{-j},y;\,q^2) \quad \text{for} \quad j=2,4,\dots,m_s-1 \quad \text{at} \quad x = 1,
     \label{eqCvee1}
\end{equation}

\begin{equation}
     \Psi_{m_s}^{C_1^{\vee}} (x q^{j},y;\,q^2) = \Psi_{m_s}^{C_1^{\vee}} (x q^{-j},y;\,q^2) \quad \text{for} \quad j=1,3,\dots,m_s \quad \text{at} \quad x = -1.
     \label{eqCvee2}
\end{equation}
The base function:
\begin{equation}
    \Psi_{m_s}^{C_1^{\vee}} (x,y;\,q^2) = x^{m_s/2 + \lambda/2} \sum_{l=0}^{m_s} \psi_l \left( \frac{1}{x} \right)^l
\end{equation}
And the solution:
\begin{equation}
    \Psi_{m_s}^{C_1^{\vee}}(x,y;\,q^2) =  \Psi_{m_s}^{B_1}(x,y;\,-q) = x^{\lambda+m_s\over 2} \sum_{k=0}^{m_s} (-q)^{k\lambda} x^{-k} \prod_{j=1}^k{[(m_s-j+1)]_{-q}[(j-\lambda-m_s-1)]_{-q}\over[j]_{-q}[(\lambda-j)]_{-q}}.
\end{equation}

\bigskip

In the next sections we move to the systems with roots with three different lengths. 

\subsection{Affine root system $BC_1$}

The Koornwinder parameters for $BC_n$ are $(a,b,c,d)=(-qt_l,-1,qt_l,t_s^2)$ 

The Koornwinder Hamiltonian:
\be
H^{K(BC_1)}_1(x):={(1-t_s^2 x )(1-q^2 t_l^2 x^2)\over \eta(1-x)(1-q^2 x^2)} \left(q^{2\hat D}-1\right)+{(1-t_s^2 x^{-1} )(1-q^2 t_l^2 x^{-2})\over \eta(1-x^{-1})(1-q^2 x^{-2})} \left(q^{-2\hat D}-1\right).
\ee
The BA function exists when 
\begin{align}
  &  t_s = q^{-m_s}, \quad m_s = 0,1,2, \dots,\\
  &  t_l = q^{-m_l}, \quad m_l = 0,2,4, \dots.
\end{align}

\begin{equation}
    \Psi^{BC_1}_{m_s,m_l} (x,y;\,q^2) = x^{\lambda+m_s/2+m_l/2} \sum_{l=0}^{m_s+m_l} \psi_l \left( \frac{1}{x}\right)^l
\end{equation}

We are left with eqs.  (\ref{eqa}), (\ref{eqc}), (\ref{eqd}):
\begin{align}
    & \Psi^{BC_1}_{m_s,m_l} (x q^{j},y;\,q^2) = \Psi^{BC_1}_{m_s,m_l} (x q^{-j},y;\,q^2) \quad \text{for} \quad j=2,4,\dots,2 m_s \quad \text{at} \quad x = 1, \\
    & \Psi^{BC_1}_{m_s,m_l} (x q^{j},y;\,q^2) = \Psi^{BC_1}_{m_s,m_l} (x q^{-j},y;\,q^2) \quad \text{for} \quad j=1,3,\dots,m_l-1 \quad \text{at} \quad x^2 = 1.
\end{align}

\subsection{Affine root system $(BC_1,C_1)$}
The Koornwinder parameters are 
\begin{equation}
    a = -t_s, \quad b =-q t_l,\quad c = t_l, \quad d=q t_l
\end{equation}
\begin{equation}
    t_s = q^{-m_s}, \quad t_l = q^{-m_l}
\end{equation}
The BA function exists when the parameters $m_s$ and $m_l$ are both odd or both even. 

The base function 
\begin{equation}
    \Psi^{(BC_1,C_1)}_{m_s,m_l}(x,y;\,q^2) = x^{\lambda+(m_s+3m_l)/4} \sum_{l=0}^{(m_s+3m_l)/2} \psi_l \left( \frac{1}{x} \right)^l
\end{equation}
satisfies the equations  
\begin{equation}
      \Psi^{(BC_1,C_1)}_{m_s,m_l}(xq^{j},y;\,q^2) =  \Psi^{(BC_1,C_1)}_{m_s,m_l}(xq^{-j},y;\,q^2),
\end{equation}
where for odd $m_s,m_l$
\begin{align}
  &  j = 1, 3, \dots m_s \quad \quad \text{at} \quad x = -1,\\
  &  j = 2, 4, \dots m_l-1 \quad \quad \text{at} \quad x^2 = 1,\\
  &  j = 1, 3, \dots m_l \quad \quad \text{at} \quad x = 1,
\end{align}
and for even $m_s,m_l$:
\begin{align}
  &  j = 1, 3, \dots m_l-1 \quad \quad \text{at} \quad x^2 = 1,\\
  &  j = 2, 4, \dots m_s \quad \quad \text{at} \quad x = -1,\\
  &  j = 2,4, \dots m_l \quad \quad \text{at} \quad x = 1.
\end{align}

\subsection{Affine root system $(C^{\vee}_1,BC_1)$}
The Koornwinder parameters are 
\begin{equation}
    a = -t_s, \quad b =-q t_s,\quad c = q, \quad d= t_l
\end{equation}
\begin{equation}
    t_s = q^{-m_s}, \quad t_l = q^{-m_l}
\end{equation}
The BA function exists when the parameter $m_l$ is even, $m_s \in \mathbb{Z}_{\geq 0}$.

The base function 
\begin{equation}
    \Psi^{(C_1^{\vee},BC_1)}_{m_s,m_l}(x,y;\,q^2) = x^{\lambda+m_s/2+m_l/4} \sum_{l=0}^{m_s+m_l/2} \psi_l \left( \frac{1}{x} \right)^l
\end{equation}
satisfies the equations  
\begin{equation}
     \Psi^{(C^{\vee}_1,BC_1)}_{m_s, m_l}(xq^{j},y;\,q^2) = \Psi^{(C^{\vee}_1,BC_1)}_{m_s, m_l}(xq^{-j},y;\,q^2), 
\end{equation}
where
\begin{align}
  &  j = 1, 2,3, \dots m_s \quad \quad \text{at} \quad x = -1,\\
  &  j = 2, 4,6, \dots m_l \quad \quad \text{at} \quad x = 1.
\end{align}

\subsection{Affine root system $(B^{\vee}_1,B_1)$}
The Koornwinder parameters are 
\begin{equation}
    a = -q, \quad b =- t_s,\quad c = q, \quad d= t_l,
\end{equation}
\begin{equation}
    t_s = q^{-m_s}, \quad t_l = q^{-m_l}.
\end{equation}
The BA function exists when both parameters $m_l$ and $m_s$ are even.

The base function 
\begin{equation}
    \Psi^{(B_1^{\vee},B_1)}_{m_s,m_l}(x,y;\,q^2) = x^{\lambda+m_s/4+m_l/4} \sum_{l=0}^{m_s/2+m_l/2} \psi_l \left( \frac{1}{x} \right)^l
\end{equation}
satisfies the equations  
\begin{equation}
     \Psi^{(B^{\vee}_1B_1)}_{m_s, m_l}(xq^{j},y;\,q^2) = \Psi^{(B^{\vee}_1B_1)}_{m_s, m_l}(xq^{-j},y;\,q^2) 
\end{equation}
where
\begin{align}
  &  j = 2,4, \dots m_s \quad \quad \text{at} \quad x = -1,\\
  &  j = 2,4, \dots m_l \quad \quad \text{at} \quad x = 1.
\end{align}

\section{Baker-Akhiezer function for root system $D_2$ \label{sec:D2case}}
The Koornwinder parameters are 
\begin{equation}
    a = -q, \quad b =- 1,\quad c = q, \quad d= 1.
\end{equation}
In case of root system $D_n$ the roots are 
\begin{equation}
    R = \left\{\pm \ve_i\pm \ve_j| 1\leq i < j \leq n\right\}, \quad\quad W \simeq S_{n} \ltimes (\mathbb{Z}/2\mathbb{Z})^{n-1},\end{equation}
    \begin{equation}
    R^+ = \left\{ \ve_i\pm \ve_j| 1\leq i < j \leq n\right\}. \label{rootsDplus}
    \end{equation}
so they are all of the same length and there is just one parameter $t = q^{-m}$. 

The base function 
\begin{equation}
    \Psi_{m}^{D_n}(\vec x,\vec y ; \,q^2) = \prod_{i=1}^n x_i^{{\lambda_i}+m(n-i)} \sum_{\{k_{ij},k'_{ij}\}=0}^{m}\psi_{\{k_{ij},k'_{ij}\}}
     \left( \prod_{1\leq i<j\leq n} \left(\frac{x_j}{x_i} \right)^{k_{ij}}\left(\frac{1}{x_i x_j} \right)^{k'_{ij}}  \right) \end{equation}
satisfies the equations
\begin{align}
     & \Psi_{m}^{D_n} (x_i q^{2s}, \vec y;\, q^2) =   \Psi_{m}^{D_n} (x_j q^{2s}, \vec y;\, q^2), \quad s = 1, \dots, m \quad \quad &&\text{at} \quad x_i = x_j, \\
    & \Psi_{m}^{D_n} (x_i q^{2s}, \vec y;\, q^2) =   \Psi_{m}^{D_n} (x_j q^{-2s}, \vec y;\, q^2), \quad s = 1, \dots, m \quad \quad &&\text{at} \quad x_i = x_j^{-1},
\end{align}

\bigskip

For the root system $D_2$ the BA function is the following
\begin{align}
  &  \Psi_m^{D_2}(x_1,x_2, \lambda_1, \lambda_2;q^2) = x_1^{\lambda_1+m}x_2^{\lambda_2} \,\,\Phi_m\left(\frac{x_1}{x_2},q^{2(\lambda_1-\lambda_2)}\right) \Phi_m\left(\vphantom{\frac{x_1}{x_2}}{x_1}{x_2},q^{2(\lambda_1+\lambda_2)}\right), 
  \label{D2expr}
\end{align}
where 
\begin{equation}
      \Phi_m(x,a;q^2)  = \sum_{k=0}^m x^{-k} \frac{(q^{-2m},q^2)_k}{(q^2,q^2)_k}\frac{(a q^{2m},q^{-2})_k}{(aq^{-2},q^{-2})_k}, 
\end{equation}

\begin{equation}
    (x,q)_k = \prod_{i=0}^{k-1}(1-x q^{i}).
\end{equation}

The answer (\ref{D2expr}) is in complete agreement with the isomorphism $D_2 \cong A_1 \times A_1$, since the BA function for $A_1$ is the following \cite{MMPapproach}:
\begin{align}
  &  \Psi_m^{A_1}(x_1,x_2, \lambda_1, \lambda_2;q^2) = x_1^{\lambda_1+m/2}x_2^{\lambda_2-m/2} \,\,\Phi_m\left(\frac{x_1}{x_2},q^{2(\lambda_1-\lambda_2)}\right).
  \label{A1expr}
\end{align}

\section{Conclusion \label{Concl}}
The paper is devoted to Chalykh's approach to the construction of the eigenfunctions of the Koornwinder Hamiltonian at the special values of parameters, i.e., when the parameters are negative integer powers of $q$. This case is connected with the affine root system $C^{\vee}C$ as well as the other affine root systems in the  $C^{\vee}C$ family. We explicitly list all the possible sets of parameters and other ingredients necessary to define the BA functions in Table \ref{tab:multirow}. We also give an explicit formula for the BA function for root system $D_2$ (\ref{D2expr}).

We closely follow the approach developed for the Ruijsenaars integrable system corresponding to the root system $A_n$. This integrable system possesses a triad of eigenfunctions (the Macdonald triad): the Noumi-Shiraishi power series (the universal solution) and its two polynomial reductions, namely Macdonald polynomials and Baker–Akhiezer functions. Koornwinder polynomials are the counterparts of symmetric Macdonald polynomials, and in this paper we consider Baker-Akhiezer functions for the root systems in the $C^{\vee}C$ family. Even though constructing the universal solution in the $C^{\vee}C$ case seems too complex, triads of eigenfunctions may still exist for the root systems in the $C^{\vee}C$ family. An important direction for future research is the construction of such analogs of the Macdonald triad, with the Baker-Akhiezer functions serving as one of their elements.

\section*{Acknowledgements}
The author is grateful to A.\,Mironov and A.\,Popolitov for stimulating discussions and valuable help in preparing this paper. 
This work is supported by the Russian Science Foundation (RSF) under grant No. 26-72-00081.

\end{document}